\documentclass[aps,pra,onecolumn,groupedaddress,showpacs]{revtex4}

\usepackage{amssymb,amsmath}
\usepackage{graphicx}

\newcommand{\om}{\omega}
\newcommand{\ve}{\varepsilon}
\newcommand{\pa}{\partial}

\begin{document}

\title{Vector $0\pi$ pulse in anisotropic media}

\author{G. T. Adamashvili}
\affiliation{Technical University of Georgia, Kostava str.77, Tbilisi, 0179, Georgia.\\ email: $guram_{-}adamashvili@ymail.com.$ }

\begin{abstract}
The system of  equations  of self-induced transparency (SIT) for extraordinary wave in uniaxial anisotropic media by means of generalized reduction perturbation method are  transformed to the coupled nonlinear Schr\"odinger equations.  It is shown that in the theory of SIT the second derivatives have significant role and leads to the formation of a vector $0\pi$ pulse oscillating with the sum and difference of the frequencies. An explicit analytical expressions for the profile and parameters of the nonlinear wave are obtained. It is shown that along with scalar $2\pi$ pulse, the  vector $0\pi$ pulse is also the basic pulse of SIT and  the scalar $0\pi$ pulse of SIT is only an approximation which can be considered in some special cases. The conditions of the existence of the nonlinear extraordinary wave depends on the direction of propagation. The profile of the vector $0\pi$ pulse in anisotropic crystal of ruby is presented with characteristic parameters which usually met in experiments.
\end{abstract}


\maketitle

\section{Introduction}

One of the most important consequence  of the light-matter interaction is formation of the nonlinear solitary waves.
The  optical resonant nonlinear waves of the stable profile can be formed when nonlinear coherent resonant interaction of an optical pulse with an  ensembles of optical  atoms or semiconductor quantum dots (SQDs) take place and the conditions of the phenomenon  of the self-induced transparency (SIT): $\omega T<<1$, $T<<T_{1,2}$  are  satisfied. Here $T$ and $\omega $ are the width and the carrier wave frequency of the optical pulse, $T_1$ and $T_2$ are the longitudinal and transverse relaxation times of the optical resonant atoms or SQDs, respectively \cite{McCall:PhysRev:69}. The intensity of the pulse interaction with atomic systems or SQDs is characterized by the area of the pulse envelope. On the other hand, the  pulse envelope area determines the type of nonlinear wave. To following the McCall-Hahn area theorem when area of the pulse envelope $\Psi$ exceed $\pi$, $2\pi$ hyperbolic secant pulse is formed and for low intensity pulses if $\Psi<<1$, than $0\pi$ pulse is generated.
The $4\pi$, $6\pi$, ... pulses in the process of propagation are divided into a discrete sequence of $2\pi$ pulses. Consequently, in resonantly absorbing media $2\pi$  and  $0\pi$ pulses are the basic pulses of SIT. They are absolutely different single-component scalar nonlinear waves with various properties and conditions of existence and have been investigated very intensively in different physical situations and various materials \cite{Allen::75, Maimistov:PhysRep:90, Panzarini::2002, Adamshvili:PhysRevA:07, Adamashvili:Phys.Rev.A:19,Crisp:PhysRep:70, Rothenberg::1984, Arkhipov::2016, Harvey::94, Arkhipov::2018, LambJr:RevModPhys:71, Poluektov:UspFizNauk:74}.

Mathematically, the interaction of an optical pulse with an ensemble of optical impurity two-level atoms is based on the Maxwell-Bloch equations.  When the pulse width is longer than a few periods of the carrier wave $ T\gg 1/\omega $, the study of SIT can be done with the slowly varying envelope approximation and reduced Maxwell-Bloch equations for the complex envelope functions of the wave and the polarization. Normally, at the solution of this system of equations  the wave equation  contains only the first-order derivatives in the space coordinate  and time and corresponding the second-order derivatives have been neglected. There are several well known methods  can use to solve SIT equations and for the investigation of the scalar single-component $ n \pi$ pulses of SIT. For instance, the factorization of the polarization and to introduce the dipole "spectral response" function \cite{McCall:PhysRev:69,Allen::75,Maimistov:PhysRep:90,LambJr:RevModPhys:71,Poluektov:UspFizNauk:74}, multiple scale method \cite{Newell::85, Dodd::1982},
the perturbative reduction method (PRM) \cite{Taniuti::1973}. The the complete solution of reduced Maxwell-Bloch equations, by means of the inverse-scattering transform (IST), are obtained \cite{Newell::85, Novikov::1984, Dodd::1982, Kaup::1977, Adamashvili:Phys.Rev.A:08}.

But the scalar single-component  nonlinear waves are not all possible SIT pulses. Recently, a completely  different type of the optical pulse of SIT compared to single-component scalar nonlinear waves  have been studied in Refs.\cite{Adamashvili:Result:11, Adamashvili:Optics and spectroscopy:2012, Adamashvili:Physica B:12, Adamashvili:Eur.Phys.J.D.:12,Adamashvili:Phys.Rev.A:14}. This is two-component (vector) pulses of SIT of low intensity with a more complex structure and different properties compared to single-component SIT pulses. Scalar single-component nonlinear waves of SIT are single waves, propagating in such a way that their parameters and shape are preserved. In contrast, the two-component vector pulse of SIT is a bound state of two nonlinear wave packets with identical polarizations and velocities. These components interact with each other and exchange energy in the process of propagation. The same polarizations of the pulse components leads to the specific profile of the resonant two-component pulse of SIT in comparison with scalar pulses of SIT and it can be considered as nonlinear pulse with specific phase modulation.
Like the scalar $0\pi$ pulse of SIT, the  pulse area of the two-component vector pulse of SIT is also zero and, therefore, it is the vector $0\pi$ pulse of SIT. Similarly to single-component pulses, the vector $0\pi$ pulse  can also propagate through the resonant medium without energy loss and it satisfies the propagation conditions of SIT.

In terms of the theory of solitons, single-component $2\pi$ and $0\pi$ pulses of SIT are scalar soliton and breather (pulsing soliton) but a resonant two-component vector $0\pi$ pulse is a vector soliton (more exactly, the nonlinear combination of the components of the vector soliton), since is a solution of the coupled nonlinear Schrodinger equations (NSEs), although it has a specific oscillation with the sum and difference of the frequencies (SDF) in the range of the carrier wave frequency (see, sections IV and V).

Influence of  the second-order derivatives  in the space coordinate  and time  of the envelope of the strength of electric field  of SIT pulse can be considered by means of different expansion methods \cite{Newell::85, Dodd::1982, Taniuti::1973} and  make sure that their contribution  leads only to small corrections to the parameters of nonlinear waves (see, for instance, \cite{Adamashvili:PRE:04, Adamashvili:PhysRevE:06}).

However, it was recently shown that taking into account in the wave equation the second-order derivatives in the space coordinate  and time we also can obtain to qualitatively new results.
It becomes possible if we consider the  SIT phenomenon and the corresponding equations  in a more general case, by applying a generalized approach [Eqs. \eqref{cemi} and \eqref{cemi1}] which is developed in references \cite{Adamashvili:Result:11, Adamashvili:Optics and spectroscopy:2012, Adamashvili:Physica B:12, Adamashvili:Eur.Phys.J.D.:12}. This is generalized version of the PRM (GPRM). Using this expansion (anzatz) it becomes possible to expand the number of auxiliary functions and parameters characterizing the wave process and get the bound state of two wave packets and as result vector $0 \pi$  pulse.

After investigating the influence of second-order derivatives in the wave equations by means of  GPRM,  an important conclusion can be made: the basic pulse of SIT along with the $2\pi$ pulse is also the vector $0\pi$ pulse, but not the scalar $0\pi$ pulse. The scalar   $0\pi$ pulse of SIT is some approximation which can be considered under the condition when we ignore the second derivatives in SIT equations or use methods (for instance, \cite{Newell::85, Dodd::1982, Taniuti::1973}) which do not allow to consider two-component waves. Consequently, for further development of the SIT theory, will be necessary to consider the properties of the vector $0\pi$ pulse  in different physical situations and in various materials.

The optical vector $0 \pi$  pulse of SIT have been investigated in different physical situations for the optical impurity atoms and SQDs. At this have been considered the pulses with different transverse structures: for the plane waves, the surface plasmon polaritons, and waveguide modes \cite{Adamashvili:Result:11, Adamashvili:Optics and spectroscopy:2012, Adamashvili:Physica B:12, Janutka::2009, Kivshar-Agrawal::03, Adamashvili:Eur.Phys.J.D.:12, Zhigang Chen::2012,  Adamashvili:PhysLettA:2015}. In all of these cases have been considered only isotropic materials in which properties of the vector $0 \pi$  pulse did not depends on the direction of pulse propagation.

On the other hand many laser crystals and nanostructures (metamaterials, two-dimensional systems) are anisotropic \cite{Kaminskii::1975}. The left-handed metamaterials are usually anisotropic and rather difficult to make isotropic one \cite{HuChui:02}. Last decade arise great interest to the anisotropic two-dimensional materials such as graphene with honeycomb hexagonal lattice, phosphorene with a puckered honeycomb configuration and strong in-plane  optical anisotropy, hexagonal boron nitride $(h-BN)$, $MoSe_{2}$, $WSe_{2}$ and others \cite{Adamashvili:Phys.Rev.A:19,Geim:Nat.Mater:07,Adamash:TPL:18}. These materials have a single axis of high symmetry of the third, fourth or sixth order. Furthermore isotropic materials can become optically anisotropic  under the influence of a deformation or  the external constant electric field. Especially interesting are properties of the uniaxial anisotropic materials in which can propagate extraordinary waves for which the energy flow  in general is not collinear to the wave vector and their futures depends on the direction of the pulse propagation \cite{Landau:Electrodynamics :84}.

The first experimental observations of the optical McCall-Hahn's  $2\pi$ and $0\pi$ scalar pulses of SIT have been demonstrated precisely in the anisotropic uniaxial ruby crystal \cite{McCall:PhysRev:69,Diels:PhysRev:74}.
Although theoretically the optical scalar single-component solitons and breathers of SIT have been investigated in anisotropic materials \cite{Agranovich::81, Sazonov::05, Maimistov::99, Adamashvili:PRE:04, Adamashvili:PhysRevE:06}, the formation of the optical vector $0 \pi$  pulse of SIT oscillating with the SDF for extraordinary waves in uniaxial anisotropic media and the conditions of the existence of this nonlinear wave have not been considered up to now.
The purpose of the present work is to  consider these problems.

\vskip+0.2cm

\section{Basic SIT equations for extraordinary wave}

We consider the optical plane extraordinary wave with width $T$, the carrier frequency $\omega$ and the wave number $k$ propagating  along the $\eta$ axis make angle $\varphi$	 with the $z$ axis, where $\eta=z \;cos \varphi +y \;sin \varphi$. We study model of the optically uniaxial media to which belong the crystals and nanostructures of the trigonal, tetragonal, and hexagonal symmetry which containing small concentration of the optical resonant impurity atoms or SQDs, $n_{0}$.  We suppose that the axis of the symmetry of third, fourth and sixth order are coincide with the $z$ axis which  are directed along with the optical axis of the uniaxial materials $O$. The principal value of the permittivity tensor $\varepsilon_{ik}$ along optical axis we determine as $\varepsilon_{zz}$. Two other principal axes are perpendicular with optical axis and are determined as $\varepsilon_{xx}$ and $\varepsilon_{yy}$. In uniaxial materials take place  the conditions  $\varepsilon_{xx}=\varepsilon_{yy}$ and  $\varepsilon_{zz}\neq\varepsilon_{xx}$. We suppose that vector of the strength of the electric field $\vec{E}$,  wave vector  $\vec{k}$, the optical axis $O$ and the axis $\eta$  lie in the same $yz$ plane. The vector of the strength of the magnetic field $\vec{H}$ is directed along the axis $x$ perpendicular to the plane of the figure \cite{Landau:Electrodynamics :84}.

The wave equation for the $z$ component of the strength of the electrical field $E_{z}$ of the optical pulse in uniaxial materials has the following form:
\begin{equation}\label{weq}
\frac{\pa^{2} E_{z}}{\pa t^2} -V^{2}(\varphi) \frac{\pa^{2} E_{z}}{\pa {\eta}^{2}}= -\frac{4\pi}{\ve_{zz}}\frac{\pa^{2}{P}}{\pa t^2} +
4\pi c^{2} \frac{cos^{2}{\varphi}}{ \ve_{zz} \;\ve_{xx} }
\frac{\pa^{2} P}{\pa {\eta}^{2}},
\end{equation}
where the quantity
\begin{equation}\label{v}\nonumber\\
V^{2}(\varphi)=c^{2}\frac{\varepsilon_{xx}+(\varepsilon_{zz}-
\varepsilon_{xx})\;cos^{2}{\varphi}}{{\varepsilon_{xx}}\;\varepsilon_{zz}},
\end{equation}
$c$ is the light velocity in vacuum. The function $P$ is the resonant nonlinear polarization due to the interaction of the optical pulse with optical impurity atoms or SQDs.

We consider pulse with the carrier frequency $\omega >>T^{-1}$  for which we can simplify Eq.\eqref{weq}  using the method of slowly changing profiles. For this purpose, we represent the function $E_{z}$  and polarization $P$ in the forms \cite{Allen::75}
\begin{equation}\label{eslow}
E_{z}(\eta,t)=\sum_{l=\pm1}\hat{E}_{l}(\eta,t)
Z_l,
\end{equation}
\begin{equation}\label{pol}
P= n_{0} d_{0} \sum_{l=\pm1} p_{l} Z_{l},
\end{equation}
where $\hat{E}_{l}$ and  ${p}_{l}$   are the slowly varying complex amplitudes of the  electric field of the optical extraordinary wave and polarization of the optical impurity atoms or SQDs, $Z_l=e^{{il(k\eta -\om t)}}$ is the fast oscillating part of the pulse amplitudes.
We shall assume, as is true of a large class of laser crystals and nanostructures, that the vector of electric dipole moment $\vec{d}_{0}$ of the optical active impurity atoms or SQDs and the optical axis $O$ of uniaxial anisotropic matrix coincide and  point along axis $z$ \cite{Adamshvili:PhysRevA:07, Kaminskii::1975, Agranovich::81}. The function $E_{z}$ is a real and therefore we assume that $\hat{E}_{l}=\hat{E}_{-l}^{\ast}$.
The quantities $p_{l}$ for optical resonant impurity atoms are determined from the optical Bloch equations, and for SQDs in the presence of single-excitonic and bi-excitonic transitions are determined from the Liouville equations \cite{Allen::75, Adamshvili:PhysRevA:07, Adamashvili:OptLett:06}.

The complex envelopes for the strength of the electric field of the pulse $\hat{E}_{l}$ vary sufficiently slowly in space and time compared with the carrier wave parts, i.e.,
\begin{equation}\label{swa}
 \left|\frac{\partial \hat{E}_{l}}{\partial t}\right|\ll\omega
|\hat{E}_{l}|,\;\;\;\left|\frac{\partial \hat{E}_{l}}{\partial \eta
}\right|\ll k|\hat{E}_{l}|.
\end{equation}
and similar expressions for the complex envelopes of polarization $p_{l}$ are valid.

Substituting Eqs.\eqref{eslow} and \eqref{pol} into Eq.\eqref{weq} we obtain the wave equation for slowly envelope amplitudes $\hat{E}_{l}$ and  ${p}_{l}$ in the form
\begin{equation}\label{eq.e}
\sum_{l=\pm1}Z_l (-2il\omega \frac{\partial \hat{E}_{l}}{\partial t} - 2ilkV^{2} \frac{\partial \hat{E}_{l}}{\partial
\eta}+\frac{\partial^{2} \hat{E}_{l}}{\partial t^{2}} -V^{2} \frac{\partial^{2} \hat{E}_{l}}{\partial \eta^{2}})=
\frac{4\pi n_{0} d_{0} {\omega}^{2}}{\varepsilon_{zz}} (1 - \frac{ c^{2} k^{2}}{{\omega}^{2}{\varepsilon_{xx} }}cos^{2}{\varphi}) \sum_{l=\pm1}p_{l}Z_l,
\end{equation}
and dispersion relation for extraordinary optical waves
\begin{equation}\label{dis}
{\omega}^{2}=V^{2}k^{2}.
\end{equation}

Following Eq.\eqref{swa} for slowly envelope amplitudes $\hat{E}_{l}$, in the theory of nonlinear waves and, in particularly, in the theory of SIT, usually it is sufficient to take into account only the first-order derivative terms of $\hat{E}_{l}$ in the space coordinate $\frac{\partial \hat{E}_{l}}{\partial \eta}$ and time  $\frac{\partial \hat{E}_{l}}{\partial t}$. The corresponding second-order derivatives $\frac{\partial^{2} \hat{E}_{l}}{\partial \eta^{2}}$ and $\frac{\partial^{2} \hat{E}_{l}}{\partial t^{2}}$   in the nonlinear wave equation \eqref{eq.e} usually have been neglected. In the frame of such approximation have been considered scalar $2\pi$  pulse (soliton) and scalar $0\pi$  pulse (breather)  solutions of the wave equation in a lot of physical situations and various materials as in isotropic, so in anisotropic media as well. Such approach have been widely used starting from the first study of the theory of SIT \cite{McCall:PhysRev:69, Allen::75, Maimistov:PhysRep:90, Panzarini::2002, Poluektov:UspFizNauk:74, Novikov::1984, LambJr:RevModPhys:71,Kaup::1977, Adamshvili:PhysRevA:07, Adamashvili:Phys.Rev.A:08, Newell::85}.

But when we neglect the second-order derivative terms  $\frac{\partial^{2} \hat{E}_{l}}{\partial \eta^{2}}$ and $\frac{\partial^{2} \hat{E}_{l}}{\partial t^{2}}$  in the nonlinear wave equation \eqref{eq.e}, at the same time arise the question: what kind of effects we ignore when we neglect the second-order derivative terms in Eq.\eqref{eq.e} and can we obtain the qualitatively new physical results when we  take into account these terms. In order to answer of this question  is necessary besides of the first-order derivative terms $\frac{\partial \hat{E}_{l}}{\partial t}$ and $\frac{\partial \hat{E}_{l}}{\partial \eta}$ must also be saved the second-order derivatives in Eq.\eqref{eq.e} and analyze this equation in more general case.

For this purpose we introduce the function of the area of the optical pulse  envelope for extraordinary optical wave
$$
\Psi_l (\eta,t)=\frac{2 d_{0}}{\hbar}\int_{-\infty}^t \hat{E_l}(\eta,t')dt',
$$
after that, the wave equation \eqref{eq.e} can be transformed  to the form
\begin{equation}\label{en. theta}
\sum_{l=\pm1}Z_l (-2il\omega \frac{\partial^{2}\Psi_{l} }{\partial
t^{2}}- 2ilkV^{2} \frac{\partial^{2} \Psi_{l} }{{\partial
\eta}{\partial t}}+\frac{\partial^{3}\Psi_{l} }{\partial t^{3}}
-V^{2} \frac{\partial^{3}\Psi_{l} }{{\partial \eta^{2}}{\partial
t}})=  \frac{8\pi{\omega}^{2}n_{0} d^{2}_{0} }{\varepsilon_{zz} \hbar} (1 - \frac{ c^{2}
k^{2}}{{\omega}^{2}{ \varepsilon_{xx} }}cos^{2}{\varphi}) \sum_{l=\pm1}p_{l}Z_l,
\end{equation}
where $\hbar$ is Planck's constant.

Eq.\eqref{en. theta}, together with  optical Bloch equations for the optical impurity atoms \cite{Allen::75} and the Liouville equations for SQDs \cite{Adamshvili:PhysRevA:07, Adamashvili:OptLett:06} are system of equations of SIT in anisotropic uniaxial media.
These systems of equations have different solutions. For small intensity pulse $|\Psi_l |<<1$ the system of the wave equation Eq.\eqref{en. theta} and optical Bloch equations (the Liouville equations) can be analyzed by means of the standard PRM \cite{Taniuti::1973} or other similar methods \cite{Newell::85, Dodd::1982, Adamashvili:PRE:04, Adamashvili:PhysRevE:04}, which allow to transform SIT equations to the NSE and obtain well known single-component breather solution-scalar $0\pi$  pulse of SIT \cite{Adamashvili:PhysRevE:06, Adamshvili:PhysRevA:07, Adamashvili:OptLett:06}.

\vskip+0.01cm
\section{GPRM and the coupled NSE$s$}

For more general analyze of the equations of SIT we make use the GPRM developed in the references \cite{Adamashvili:Result:11, Adamashvili:Optics and spectroscopy:2012, Adamashvili:Physica B:12, Adamashvili:Eur.Phys.J.D.:12, Adamashvili:PhysLettA:2015}.
This method give possibility to transform some nonlinear equations to the coupled NSEs and it provides the solution of these equations in the form of the two-component vector $0\pi$ pulse oscillating with the SDF. Using this approach the complex function $\Psi_{l}(\eta,t)$ in Eq.\eqref{en. theta}   can be represented as:
\begin{equation}\label{cemi}
\Psi_{l}(\eta,t)= \sum_{\alpha=1} \varepsilon^\alpha {{\Psi}_{l}}^{(\alpha)}(\eta,t),
\end{equation}
where
\begin{equation}\label{cemi1}
{{\Psi}_{l}}^{(\alpha)}(\eta,t)= \sum_{n=-\infty}^{+\infty} Y_{l,n} f_{l,n}^ {(\alpha)}(\zeta_{l,n},\tau),
\end{equation}
$$
Y_{l,n}=e^{in(Q_{l,n}\eta-\Omega_{l,n}
t)},
\;\;\;\;\;\;\;\;\;\;\;
\zeta_{l,n}=\varepsilon Q_{l,n}(\eta-v_{l,n}
t),
\;\;\;\;\;\;\;\;\;
\tau=\varepsilon^2 t,
\;\;\;\;\;\;\;\;\;
{v}_{l,n}=\frac{d\Omega_{l,n}}{dQ_{l,n}},
$$
$\varepsilon$ is a small parameter. Such a representation allows us to separate from $\Psi_{l}$ the still more slowly changing
auxiliary functions $ f_{l,n}^{(\alpha )}$. Consequently, it is assumed that the quantities $\Omega_{l,n}$, $Q_{l,n}$ and $f_{l,n}^{(\alpha)}$ satisfies the inequalities for any $l$ and $n$:
\begin{equation}\label{rtyp}\nonumber\\
\omega\gg \Omega_{l,n},\;\;k\gg Q_{l,n},\;\;\;
\end{equation}
$$
\left|\frac{\partial
f_{l,n}^{(\alpha )}}{
\partial t}\right|\ll \Omega_{l,n} \left|f_{l,n}^{(\alpha)}\right|,\;\;\left|\frac{\partial
f_{l,n}^{(\alpha )}}{\partial \eta }\right|\ll Q_{l,n} \left|f_{l,n}^{(\alpha )}\right|.
$$

Unlike the standard PRM \cite{Taniuti::1973}, in the expansions Eqs.\eqref{cemi} and \eqref{cemi1}, the quantities $\Omega_{l,n}$, $Q_{l,n}$, $v_{l,n}$, $\zeta_{l,n}$ and $f_{l,n}^{(\alpha)}$  are depends from the indexes $l$ and $n$.
In the special case when the quantities  $\Omega_{l,n}$, $Q_{l,n}$, $v_{l,n}$, $\zeta_{l,n}$ and $f_{l,n}^{(\alpha)}$  are not depends from the indexes $l$ and $n$ the expansions Eqs.\eqref{cemi} and \eqref{cemi1}  are transformed to the well known standard PRM \cite{Taniuti::1973}.
\begin{equation}\label{taniuta}
{{\Psi}}(\eta,t)= \sum_{n=-\infty}^{+\infty}\varepsilon^{\alpha} e^{in(Q\eta-\Omega t)} f_{n}^ {(\alpha)}(\zeta,\tau).
\end{equation}
It should be noted that, in contrast to PRM Eq.\eqref{taniuta} \cite{Taniuti::1973}, in which only auxiliary function $f_{\pm 1}^ {(\alpha)}$ and two parameters $\Omega$ and $Q$ appear, in the generalized PRM Eqs.\eqref{cemi} and \eqref{cemi1}, auxiliary functions $f_{\pm 1, \pm 1}^ {(\alpha)}$ and $f_{\pm 1, \mp 1}^ {(\alpha)}$ and parameters $\Omega_{\pm 1,\pm 1}$, $\;\Omega_{\pm 1,\mp 1}$, $Q_{\pm 1,\pm 1}$ and $Q_{\pm 1,\mp 1}$ are used. It is through more auxiliary functions and parameters of the GRPM that a more general and much more detailed description of the wave process becomes possible.

In further, for simplicity, we omit $l$ and $n$ indexes for the quantities $\Omega_{l,n}$, $Q_{l,n}$, $v_{l,n}$ and $\zeta_{l,n}$ in equations where this will not bring about mess.

Substituting Eq.\eqref{cemi} into Eq.\eqref{en. theta} we obtain the equation
\begin{equation}\label{wqq}
\sum_{l=\pm1}\sum_{\alpha=1} \varepsilon^\alpha  Z_l (-2il\omega
\frac{\partial^{2}{\Psi}^{(\alpha)}_{l} }{\partial t^{2}}-
2ilk V^{2} \frac{\partial^{2} {\Psi}^{(\alpha)}_{l}  }{{\partial
\eta}{\partial t}}+\frac{\partial^{3}{\Psi}^{(\alpha)}_{l}
}{\partial t^{3}} -V^{2} \frac{\partial^{3}{\Psi}^{(\alpha)}_{l}
}{{\partial \eta^{2}}{\partial t}})=$$$$i \beta^{2}\sum_{l=\pm 1}l
Z_l (\varepsilon^{1} {{\Psi}^{(1)}_{l}}
 +\varepsilon^{2} {{\Psi}^{(2)}_{l}}+\varepsilon^{3}{{\Psi}^{(3)}_{l}}
 -\varepsilon^{3}
\frac{1}{2} \int\frac{\pa {{\Psi}^{(1)}_{l}}}{\pa
t}{{\Psi}^{(1)}_{-l}} {{\Psi}^{(1)}_{l}} dt')
 +O(\ve^4),
\end{equation}
where
\begin{equation}\label{beta}
\beta^{2}(\varphi)=\frac{4 \pi {\omega}^{2}n_{0}{d_{0}}^{2}
 }{\hbar \;\varepsilon_{zz}} (1 - \frac{ c^{2} k^{2}}{{\omega}^{2}{
\varepsilon_{xx} }}cos^{2}{\varphi}) \int \frac{g(\Delta)}{1+\Delta^{2}T^{2}}d\Delta,
\end{equation}
$g(\Delta)$ is the inhomogeneous broadening lineshape function for an ensemble of optical resonance atoms or SQDs, $\Delta=\omega_{0}-\omega,\;\;\om_{0}$ is the transition frequency of the optical resonant atoms (SQDs).

From the Eq.\eqref{beta} we can see that nonlinear extraordinary wave can not be formed for any direction of propagation. Indeed, for  small values of $\varphi$ the coefficient  $\beta^{2}(\varphi)$, which characterized the strength  of the light-atoms (SQDs) interaction, is also very small. In such directions nonlinear effects will be expressed very weakly and it is not enough for the formation of the nonlinear wave (see criterion in Eq.\eqref{kr}).

Substituting the expansion  \eqref{cemi1}  into  \eqref{wqq} we obtain wave equation in uniaxial anisotropic media
\begin{equation}\label{ww}
\sum_{l=\pm1}\sum_{\alpha=1}\sum_{n=-\infty}^{+\infty}
  \varepsilon^\alpha  Z_{l} Y_{l,n}\{\mathfrak{W}_{l,n} +\varepsilon \mathfrak{J}_{l,n}\frac{\partial }{\partial \zeta}
 +\varepsilon^2 \mathfrak{h}_{l,n} \frac{\partial }{\partial \tau}+
\varepsilon^{2}i \mathfrak{H}_{l,n} \frac{\partial^{2} }{\partial
\zeta^{2}}\}f_{l,n}^ {(\alpha)}$$$$=-\varepsilon^{3}\frac{i}{2} \beta^{2}\sum_{l=\pm 1}l Z_l
 \int\frac{\pa {{\Psi}_{l}}^{(1)}}{\pa
t}{{\Psi}_{-l}}^{(1)} {{\Psi}_{l}}^{(1)}dt'
 +O(\ve^4),
\end{equation}
where
\begin{equation}\label{www}
\mathfrak{W}_{l,n}=in \Omega [A_{l}n {\Omega} - B_{l} n Q  +
{\Omega}^{2}-V^{2}  Q^{2}  -\frac{l}{n }\frac{\beta^{2}}{ \Omega}],
$$$$
\mathfrak{J}_{l,n}= n Q[2A_{l} \Omega v -B_{l}(Q v +\Omega) + 3n {\Omega}^{2}  v -V^{2}n Q(Q v + 2 \Omega)],
$$$$
\mathfrak{h}_{l,n}=-2 n A_{l}\Omega + B_{l}n Q - 3  {\Omega}^{2} +V^{2}   Q^{2},
$$$$
\mathfrak{H}_{l,n}=  Q^{2} [-A_{l}   v^{2} + B_{l}   v  -3n\Omega  {v}^{2} + V^{2}n ( 2Q v +\Omega) ],
$$$$
A_{l}=2l\omega,\;\;\;\;\;\;\;\;\;\;\;\;\;\;\;\;B_{l}=2lk V^{2}.
\end{equation}

Following the standard procedure characterized for any perturbative expansions and equate to each other the terms of the same order to $\varepsilon$, from the Eqs.\eqref{ww} and \eqref{www} we obtain the chain of the equations. As result we determine the connection between the parameters $\Omega_{l,n}$  and $Q_{l,n}$ which is given by
\begin{equation}\label{dis2}
l n (2 \omega \Omega_{l,n} -2 k V^{2}  Q_{l,n} -  \frac{\beta^{2}}{\Omega_{l,n}}) = V^{2} Q^{2}_{l,n} -  \Omega^{2}_{l,n}
\end{equation}
and the coupled NSEs in the following form
\begin{equation}\label{cnse2}
 i  (\frac{\partial U_{\pm}}{\partial t}+ v_{\pm} \frac{\partial U_{\pm}}{\partial \eta})
+  p_{\pm} \frac{\partial^{2} U_{\pm}
}{\partial \eta^{2}} + \mathfrak{q}_{\pm} |U_{\pm}|^{2}U_{\pm} +r_{\pm} |U_{\mp}|^{2} U_{\pm}=0,
\end{equation}
where
\begin{equation}\label{koef}
U_{\pm}=\ve f_{+1,\pm 1}^ {(1)},\;\;\;\;\;\;\;\;\;\;\;\;\;\;v_{\pm}=v_{+1,\pm 1},\;\;\;\;\;\;\;\;\;\;\;
p_{\pm}=-\frac{\mathfrak{H}_{+1,\pm 1}}{\mathfrak{h}_{+1,\pm 1}Q^{2}_{\pm 1}},
$$$$
\mathfrak{q}_{\pm}=-\frac{ \beta^{2}}{2 \mathfrak{h}_{+1,\pm 1}},\;\;\;\;\;\;\;\;\;\;\;\;\;\;\;\;
r_{\pm}=q_{\pm}(1  - \frac{\Omega_{\mp 1} }{\Omega_{\pm 1}}).
\end{equation}

\vskip+0.01cm
\section{Optical vector $0\pi$ pulse of SIT}

The solutions of the Eq.\eqref{dis2} depend  on the variables $l=\pm 1$ and $n=\pm 1$. Therefore we have four solutions $Q_{+1,+1}$,  $Q_{+1,-1}$,  $Q_{-1,+1}$ and  $Q_{-1,-1}$ for corresponding values of the quantities $\Omega_{l,n}$. But because  Eq.\eqref{dis2} depends only on the products of variables $l\;n$
we obtain the equations
\begin{equation}\label{koef1}
\Omega_{+1,-1}=\Omega_{-1,+1}=\Omega_{-1},\;\;\;\;\;\;\Omega_{+1,+1}=\Omega_{-1,-1}=\Omega_{+1},
$$
$$
Q_{+1,-1}=Q_{-1,+1}=Q_{-1},\;\;\;\;\;\;Q_{+1,+1}=Q_{-1,-1}=Q_{+1}.
\end{equation}

From Eqs. \eqref{dis2}, \eqref{koef} and \eqref{koef1} are obvious that the all these parameters depends from the direction of the pulse propagation (angle $\varphi$).

The terms on the right hand side of the  Eq.\eqref{dis2}    $V^{2} Q^{2}$ and $ \Omega^{2} $ arise in this equation from the second-order derivatives of the wave equation Eq.\eqref{eq.e}  $V^{2} \frac{\partial^{2} \hat{E}_{l}}{\partial \eta^{2}}$  and $\frac{\partial^{2} \hat{E}_{l}}{\partial t^{2}}$.
If we neglect these second-order derivatives in Eq.\eqref{eq.e}, as usually have made in the theory of SIT \cite{McCall:PhysRev:69, Allen::75, Maimistov:PhysRep:90, Panzarini::2002, Poluektov:UspFizNauk:74, LambJr:RevModPhys:71, Kaup::1977, Adamashvili:Phys.Rev.A:08, Newell::85}, the right hand side of the  Eq.\eqref{dis2} will be equal to zero and the parameters $\Omega$  and $Q$ will not depends from the indexes $l$ and $n$.
Consequently, under this condition we obtain that $\Omega_{+1}=\Omega_{-1}=\Omega,\;$ $\;r_{\pm}=0$ and the coupled NSEs Eq.\eqref{cnse2} will be disconnected to the two independent NSEs. As result, we obtain two independent single-component scalar $0\pi$ pulses (breathers) which will be propagating separately to each other.

From this we can make the conclusion that the second-order derivatives in the space coordinate and time in the wave equation \eqref{eq.e} give effect of the interaction between two breathers and provide the formation of a bound state of these breathers, i.e. the small intensity two-component vector $0\pi$ pulse.

It should be noted hat the analyse of Eq.\eqref{en. theta} by means of expansion \eqref{cemi} is valid only in case when  the area of pulse $\Psi_{l}(\eta,t)$ is complex function, i.e. when the phase modulation of the nonlinear pulse take place. Otherwise, when the phase modulation is absent, the functions $\Psi_{l}(\eta,t)$  and $\hat{E}_{l}(\eta,t)$ are transformed to the real functions  $\Psi(\eta,t)$  and $\hat{E}(\eta,t)$ which are independent of the index  $l$. Under these conditions the wave equation \eqref{en. theta} impossible to transform to the coupled NSEs (exception for real function, see Ref.\cite{Adamashvili:Result:11}).

From Eq.\eqref{dis2} we obtain the expression for the parameter
\begin{equation}\label{v}
v_{\pm }=\frac{\pm  k +   Q_{\pm 1} }{\pm  \omega  +   \Omega_{\pm 1}  \pm \frac{\beta^{2}}{2 \Omega^{2}_{\pm 1}}}V^{2}.
\end{equation}

The coupled NSEs \eqref{cnse2} have been analyzed widely many years (see, for instance \cite{Adamashvili:Optics and spectroscopy:2012, Adamashvili:Eur.Phys.J.D.:12} and references therein). In the simplest case when all coefficients are equal to each other, the set of two equations \eqref{cnse2} are solvable by IST \cite{Manakov::74}.

We seek solutions of Eqs.\eqref{cnse2} in the form of
\begin{equation}\label{ue}
U_{\pm}(\mu,t)=K_{\pm 1}\mathfrak{S}(\mu)e^{i(k_{\pm 1} \eta - \omega_{\pm 1} t )},
\end{equation}
where the quantities $K_{\pm 1},\; k_{\pm 1}$ and $\omega_{\pm 1}$ are the real constants.
$\mu=t-\frac{\eta}{V_{0}},\;$ $V_{0}$ -the velocity of the nonlinear wave. The following inequalities
\begin{equation}\label{kom}
k_{\pm 1}<<Q_{\pm 1},\;\;\;\;\;\;\omega_{\pm 1}<<\Omega_{\pm 1},
\end{equation}
are fulfilled.

Substituting Eq.\eqref{ue} into Eqs.\eqref{cnse2} we obtain
$$
\mathfrak{S}(\mu)=\frac{1}{\mathfrak{b} T}Sech(\frac{\mu}{T}),
$$
where the relations between quantities $K_{\pm 1}$ and $\omega_{\pm 1}$ have the forms
\begin{equation}\label{rt16}
K_{+1}^{2}=\frac{p_{+}\mathfrak{q}_{-}- p_{-}r_{+}}{p_{-}\mathfrak{q}_{+}-p_{+}r_{-}}K_{-1}^{2},
\;\;\;\;\;\;\;\;\;
\omega_{+1}=\frac{p_{+}}{p_{-}}\omega_{-1}+\frac{V^{2}_{0}(p_{-}^{2}-p_{+}^{2})+v_{-}^{2}p_{+}^{2}-v_{+}^{2}p_{-}^{2}
}{4p_{+}p_{-}^{2}}.
\end{equation}

Eq.\eqref{ue} has analytical solutions in the form of breather pairs that preserve their profile through the propagation
\begin{equation}\label{ue1}
U_{\pm}(\mu,t)=\frac{K_{\pm 1}}{\mathfrak{b} T}Sech(\frac{\mu}{T}) e^{i(k_{\pm 1} \eta - \omega_{\pm 1} t )}.
\end{equation}

Substituting Eq.\eqref{ue1} into Eqs.\eqref{cemi}, \eqref{cemi1} and \eqref{eslow} we obtain the vector $0\pi$ pulse of SIT with the SDF for extraordinary wave in uniaxial anisotropic media:
\begin{equation}\label{vs}
E_{z}(\eta,t,\varphi)= \frac{\hbar }{\mathfrak{b}(\varphi) T d_{0}}Sech(\frac{t-\frac{\eta}{V_{0}}}{T})\sum_{j=\pm 1}    \{j (\Omega_{j}(\varphi)+j \omega_{j}) K_{j }(\varphi) \sin[(k +j Q_{j }(\varphi)+k_{j})\eta -(\om +j \Omega_{j}(\varphi)+\omega_{j}) t] \}.
\end{equation}
where
\begin{equation}\label{er17}
T^{-2}=V_{0}^{2}\frac{v_{+}k_{+1}+k_{+1}^{2}p_{+}-\omega_{+1}}{p_{+}},\;\;\;
{\mathfrak{b}}^{2}=V_{0}^{2} \frac{K_{+1}^{2}\mathfrak{q}_{+}+K_{-1}^{2}r_{+}}{2p_{+}},\;\;\;\;\;k_{\pm 1}=\frac{V_{0}-v_{\pm}}{2p_{\pm}}.
\end{equation}

From Eq.\eqref{vs} is obvious that  the trigonometric functions  $\sin[(k + Q_{+1 }+k_{+1})\eta -(\om + \Omega_{+1}+\omega_{+1}) t]$ and $\sin[(k - Q_{-1 }+k_{-1})\eta -(\om - \Omega_{-1}+\omega_{-1}) t]$ points of the exitances of oscillations with the characteristic parameters at the sum $\om + \Omega_{+1} (k + Q_{+1 })$  and difference $\om - \Omega_{-1} (k - Q_{-1 })$ [to taking into account Eq.\eqref{kom}] of the frequencies (wave number) in the region of the carrier wave frequency $\omega$ and wave number $k$. This leads to the fact that solutions of the coupled NSEs for the auxiliary functions $U_{\pm}$ are transformed to the Eq.\eqref{vs} for the profile of  the phase modulated z component of the strength of the electric field $E_{z}$ of the extraordinary wave. The parameters of the resonant   vector 0$\pi$ pulse of SIT oscillating with the SDF are determined from the Eqs.\eqref{koef}, \eqref{v}, \eqref{rt16} and \eqref{er17}. The dispersion relation for extraordinary wave and connection between oscillating parameters   $\Omega_{\pm 1}$ and $Q_{\pm 1}$  are determined from equations \eqref{dis} and \eqref{dis2}.
From the Eqs.\eqref{dis2},  \eqref{koef}, \eqref{koef1} and \eqref{v}   obvious that the parameters $b$, $\Omega_{j}$, $Q_{j}, K_{j}$  of the nonlinear pulse Eq.\eqref{vs} are depends from the direction of propagation, i.e. from the angle $\varphi$.

Using typical parameters  for the small intensity vector $0\pi$ pulse of SIT with the SDF for extraordinary wave Eq.\eqref{vs}, and use the crystal that have been  applied for experimental observing of SIT \cite{McCall:PhysRev:69, Diels:PhysRev:74, Poluektov:UspFizNauk:74}. A plot of the strength of the electric field of the two-component vector $0\pi$ pulse of SIT
in crystal of ruby ($Al_{2}O_{3}:Cr^{+3}$), depending from the direction of propagation, i.e. on the angle $\varphi$, is shown in Figure 1 for a fixed value of the coordinate at  $\eta = 0$.

The parameters of the medium and pulse are:
$\omega=6 \times 10^{15}\; s^{-1},\;$ $T =3\times10^{-8}\; s,\;$  $d_{0}=5 \times 10^{-21}\; CGSE\; units,\;\;$$n_{0}= 8 \times 10^{15} \;cm^{-3},\;\;$  $T^{*} = 0.03 \;ns.\;\;$
$\varepsilon_{xx}=3.1329,\;\;$
$\varepsilon_{zz}=3.10464.$
\begin{figure}[htbp]
\includegraphics[width=0.44\textwidth]{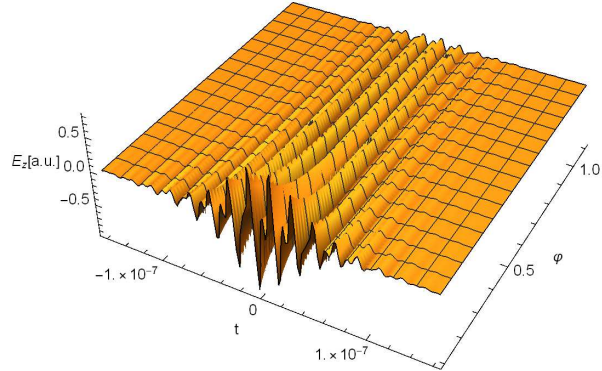}
\caption{The dependence of the $z$-component of the strength of electric field $E_{z}$  of the low intensity vector $0\pi$ pulse of SIT with the SDF for extraordinary wave Eq.\eqref{vs} from the time $t$ and direction of propagation $\varphi$,  at  $\eta=0$, in crystal of ruby is presented.}
\label{fig1}
\end{figure}

\vskip+0.01cm
\section{Conclusions and Discussions}

At the theoretical description of the coherent interaction of light with an anisotropic materials, normally proceed from the Maxwell wave equation for extraordinary wave Eq.\eqref{weq}. For pulses with duration $T >> 1 / \omega$ usually used the slowly varying envelope approximation Eqs.\eqref{eslow}, \eqref{pol} and \eqref{swa}. For slowly variables the SIT equations are the Maxwell wave equation Eq.\eqref{eq.e} which contains the first-order and second-order derivatives in the space coordinate $\eta$ and time $t$  and the system of Bloch equations for the envelope functions.  This system of equations can be solved by means of several methods and obtain all well known one-component solutions for scalar $2\pi n$ pulses of SIT.
Lamb showed that $4\pi$  , $6\pi$, ... pulses are divided into a discrete sequence of $2\pi$   pulses \cite{Allen::75}. Consequently, in resonantly absorbing media, the basic SIT pulses are:  $2\pi$  and $0\pi$  pulses. These pulses are single-component scalar nonlinear waves and these statements are valid only under the condition when in wave equation the second-order derivatives are neglected
\cite{McCall:PhysRev:69,Allen::75,LambJr:RevModPhys:71,Poluektov:UspFizNauk:74,Agranovich::81,Adamashvili:PhysRevE:04}.

However, the second-order derivatives contain important information about both single-component and two-component nonlinear waves and play a significant role in various wave processes. Using the well-known methods
we can investigate various effects associated with the second derivatives in the wave Maxwell equations only for single-component waves. For example, using the multi-scale method, we can investigate the optical self-focusing of a laser beam in dielectrics and Langmuir waves in a plasma \cite{Dodd::1982}. Using PRM, we can determine quantitative corrections to the parameters of scalar $0\pi$  pulse \cite{Adamashvili:PRE:04, Adamashvili:PhysRevE:06}.

But for the study of two-component waves, another method is necessary that contains more auxiliary functions and parameters. Precisely such  method is  GPRM [Eqs.\eqref{cemi} and \eqref{cemi1}] \cite{Adamashvili:Result:11, Adamashvili:Optics and spectroscopy:2012, Adamashvili:Physica B:12, Adamashvili:Eur.Phys.J.D.:12, Adamashvili:PhysLettA:2015}. Using this method, the nonlinear wave equation \eqref{en. theta} for extraordinary wave is transformed to the coupled NSEs \eqref{cnse2}. Substituting the solutions of the Eqs.\eqref{cnse2} into Eq.\eqref{eslow} we obtain Eq.\eqref{vs},  the phase modulated vector 0$\pi$ pulse of SIT oscillating with the SDF of the extraordinary wave in uniaxial anisotropic media.
After this study, it is obvious that the basic SIT pulses are  scalar  $2\pi$  pulse and vector  $0\pi$  pulse, but not scalar $0\pi$  pulse. The scalar  $0\pi$  pulse is only an approximation which we can consider in the case when we neglecting the second-order derivatives in the  Maxwell wave equation or using methods do not contains enough auxiliary functions and parameters for investigation two-component pulses \cite{Allen::75, Newell::85, Novikov::1984, Dodd::1982}.

Summarizing  we can formulating the conditions of the existence of the small intensity phase modulated  resonant   vector 0$\pi$ pulse of SIT oscillating with the SDF in uniaxial anisotropic media  Eq.\eqref{vs}:

 1i. Taking into account the second derivative terms of $E_{z}$ with respect to the space coordinates  $\frac{\partial^{2} \hat{E}_{l}}{\partial \eta^{2}}$ and time $\frac{\partial^{2} \hat{E}_{l}}{\partial t^{2}}$   in the nonlinear wave equation \eqref{eq.e}.

 2i. To consider the situation with the different values of the oscillating parameters $\Omega_{\pm 1}$ and $Q_{\pm 1}$  which is possible only if using the GPRM [Eqs.\eqref{cemi} and \eqref{cemi1}].

 Otherwise, if we using standard PRM Eq.\eqref{taniuta} \cite{Taniuti::1973}, we obtain that $\Omega_{+1}=\Omega_{-1}=\Omega$, $\;Q_{+1}=Q_{-1}=Q$ and $r_{\pm}=0$ and consequently instead of vector 0$\pi$ pulse of SIT Eq.\eqref{vs} we get only two uncoupled breathers propagating independently of each other. If in this case, i.e. if we use Eq.\eqref{taniuta}, and to take into account the second derivative terms, this will only leads to small corrections to the parameters of independent breathers \cite{Adamshvili:PhysRevA:07, Adamashvili:PhysRevE:06}.

 3i. To consider the phase modulation of the nonlinear pulse, i.e. it is necessary that the functions $\Psi_{l}(\eta,t)$  and $\hat{E}_{l}(\eta,t)$ must be complex functions. Otherwise, we can not using the GPRM Eqs.\eqref{cemi} and \eqref{cemi1} for the wave equation and consequently can not obtained vector 0$\pi$ pulse solution Eq.\eqref{vs}.

 4i. From the Eq.\eqref{beta} we can see that nonlinear wave can not be formed for any direction of propagation. For the small values of $\varphi$, the coefficient  $\beta^{2}(\varphi)$,  that characterizes the strength  of the light - atoms (SQDs) interaction, is also very small. For such directions nonlinear effects will be expressed very weakly and  not enough for the formation of the nonlinear wave.  In particular, along optical axes $\varphi=O$ can not be propagating vector 0$\pi$ pulse of SIT.

 When the conditions
 \begin{equation}\label{kr}
 \mathfrak{q}_{+}\sim p_{+},\;\;\;\;\;\;\;\;\;\;\;   \mathfrak{q}_{-}\sim p_{-}
 \end{equation}
are fulfilled, i.e., $ \mathfrak{q}_{+}$ and $p_{+}$, and also $ \mathfrak{q}_{-}$ and $p_{-}$ are the quantities of the same order, the vector 0$\pi$ pulse of SIT  Eq.\eqref{vs} can  be formed.
Because the quantities  $p_{\pm}$ and  $\mathfrak{q}_{\pm}$ depends from the direction of the pulse propagation (angle $\varphi$), consequently vector 0$\pi$ pulse of SIT Eq.\eqref{vs} can propagating only for such values of $\varphi$  for which Eq.\eqref{kr} are fulfilled.

Finally we can make the conclusion that the small intensity  vector $0\pi$ pulse of SIT oscillating with the SDF   for the extraordinary wave Eq.\eqref{vs} can be formed only in such physical situations when all 1i-4i requirements are fulfilled.

In the particular case, when  angle $\varphi=\pi/2$ all above presented expressions  coincide with the previously obtained results in isotropic medium \cite{Adamashvili:Optics and spectroscopy:2012}.

We study propagation of the extraordinary waves for plane waves and for optical resonance atoms but generalization obtained results for surface plasmon polaritons or for waveguide modes in multi-layered uniaxial anisotropic materials (metamaterials, two-dimensional systems) and for SQDs is possible without much difficulty.

This work allow to obtained more comprehensive theory of SIT for extraordinary waves in uniaxial anisotropic media. The previously studied  the single-component  $2\pi$ and $0\pi$ scalar pulses (soliton and breather) \cite{Agranovich::81, Sazonov::05, Maimistov::99, Adamashvili:PRE:04, Adamashvili:PhysRevE:06}, is supplemented by the  two-component vector 0$\pi$ pulse which is one of the basic pulse of SIT.

The above reported results give grounds to hope that the vector 0$\pi$ pulse of SIT oscillating with the SDF  can be observed experimentally.

\vskip+0.01cm
\section{Acknowledgments}

G.T.A. acknowledges the Shota Rustavely NSF Grant No. 217064 for the support of this work.

\vskip+0.5cm

\end{document}